\documentstyle[11pt,newpasp,twoside,epsf,graphicx]{article}
\def\edcomment#1{\iffalse\marginpar{\raggedright\sl#1\/}\else\relax\fi}
\reversemarginpar

\begin{document}
\title{Coronae of Cool Stars}
\author{M.~Audard$^1$, S.~A.~Drake$^{2,3}$, M.~G\"udel$^4$, R.~Mewe$^5$,
R.~Pallavicini$^6$, T.~Simon$^7$, K.~P.~Singh$^8$, S.~L.~Skinner$^9$,
N.~E.~White$^3$}

\affil{$^1$Columbia University, $^2$USRA, $^3$NASA/GSFC, $^4$Paul Scherrer
Institut, $^5$SRON-Utrecht, $^6$Palermo Observatory, $^7$University of Hawaii,
$^8$Tata Institute of Fundamental Research, $^9$CASA/Univ. of Colorado}

%

%
%
%
%
%
%

\begin{abstract}
We present preliminary results of grating observations of YY Mensae and V824
Arae by Chandra and XMM-Newton. Spectral features are presented in
the context of the emission measure distributions, the coronal abundances, and
plasma electron densities. In particular, we observe a coronal N/C 
enhancement in YY Men believed to reflect the photospheric composition
(CN cycle). Finally, we interpret line broadening in YY 
Men as Doppler thermal broadening in its very hot corona.
\end{abstract}

\vspace*{-10mm}

\section{Introduction}
Grating spectrometers on board Chandra and XMM-Newton have revolutionized the
field of X-ray spectroscopy of cosmic plasmas. Individual lines can be
measured in the soft X-ray energy range with sufficient accuracy to permit
measurements of elemental abundances, plasma densities, or simply to reveal the
spectroscopic nature of a cosmic plasma (e.g., see Paerels \& Kahn 2003 for a
review on high-resolution X-ray spectroscopy with Chandra and XMM-Newton).

Many cool stars are bright X-ray sources thanks to their hot coronae and their close
distances. They are ideal targets for X-ray spectroscopy since they display a
wealth of bright emission lines of various elements. Recent reviews summarize
results obtained in the first years after the launch of Chandra and XMM-Newton
(e.g., Audard 2003, G\"udel 2003, Linsky 2003). In particular, patterns in
coronal abundances have been recognized in which elements with a first ionization
potential (FIP) $<10$~eV are underabundant with respect to the high-FIP
elements when observed in very active stars; this is opposite to the solar FIP
effect in which low-FIP elements are overabundant whereas high-FIP elements are
of photospheric composition. However, coronal abundances in a sample of stars with 
a broad range of magnetic activity seem to reverse from a solar-like
FIP effect in the least active, old stars to an ``inverse''-FIP effect in the
most active, young stars (G\"udel et al.~2002). 
The trend is suggestive and a larger sample is certainly needed.

Two very active stars, YY Men and V824 Ara, have been observed with
grating spectrometers on board Chandra and XMM-Newton in order to study, in
particular, their coronal abundances. YY Men is one of the (rare) class of
single giants with very hot coronae, whereas V824 Ara is an example of a close
main-sequence, very active RS CVn binary.
We present here preliminary results of the grating observations.

\section{Observations and Data Reduction}

Chandra observed YY Men in the ACIS-S/HETG configuration for 74~ks in
February 1--2, 2002, whereas the 86~ks-long XMM-Newton observation dates from October 5--6,
2001. The RS CVn binary V824 Ara was observed with Chandra 
ACIS-S/HETGS for 94~ks on July 10--11, 2002. 
The Chandra data sets were reduced with CIAO 2.3 with CALDB 2.21 using standard
techniques described in threads; however, events for V824 Ara and LDS 587B were 
resolved using narrower, non-standard grating arms (10 times the 1$\sigma$ 
cross-dispersion width instead of the default 35 times). Thanks to the low 
background in ACIS, no background spectra were used in the Chandra spectral 
analysis allowing us to use the robust C statistics. 
The XMM-Newton grating data were reduced with SAS 5.4.1 using standard
techniques (see Audard et al.~2003).

\vspace*{-4mm}
\section{YY Mensae = HD 32918}
The Chandra and XMM-Newton observations were motivated by the very hot
temperature detected with ROSAT PSPC and ASCA (G\"udel et al.~1996) and its 
extreme luminosity ($\log L_{\rm X} \sim 32$ for a distance of 291~pc). 
YY Men ($P \sim 9.55$~d; $v \sin i \sim 50$~km~s$^{-1}$; e.g., Collier Cameron 
1982) is a member of a loosely-defined class of rapidly rotating single
G and K giant stars (``FK Com stars''), whose outstanding property is a projected 
equatorial velocity measured up to 160~km~s$^{-1}$, in contrast with the 
expected maximum of 6~km~s$^{-1}$ for giants. One of the leading theories to 
explain the extreme properties of FK Com stars suggests that they were
formed by coalescence of a contact binary when one of the components entered 
into the giant stage (Bopp \& Stencel 1981).

Figure~1 shows the Chandra ACIS-S/HETG first-order HEG and MEG spectra. Several
emission lines are detected; in particular Fe lines in various ionization states
can be found, from Fe~{\sc xvii} to Fe~{\sc xxv}, indicating a broad
distribution of plasma temperatures, typically from $4$~MK to $30$~MK. Elements
such as N, O, Ne, Mg, Si, S, Ar, and Ca are detected as well, mostly in
H-like and He-like transitions. The strong Ly$\alpha$ lines indicated a dominant
very hot plasma, also suggested by the Fe K$\alpha$ complex and the strong
underlying continuum.

\begin{figure}[!t]
\plotone{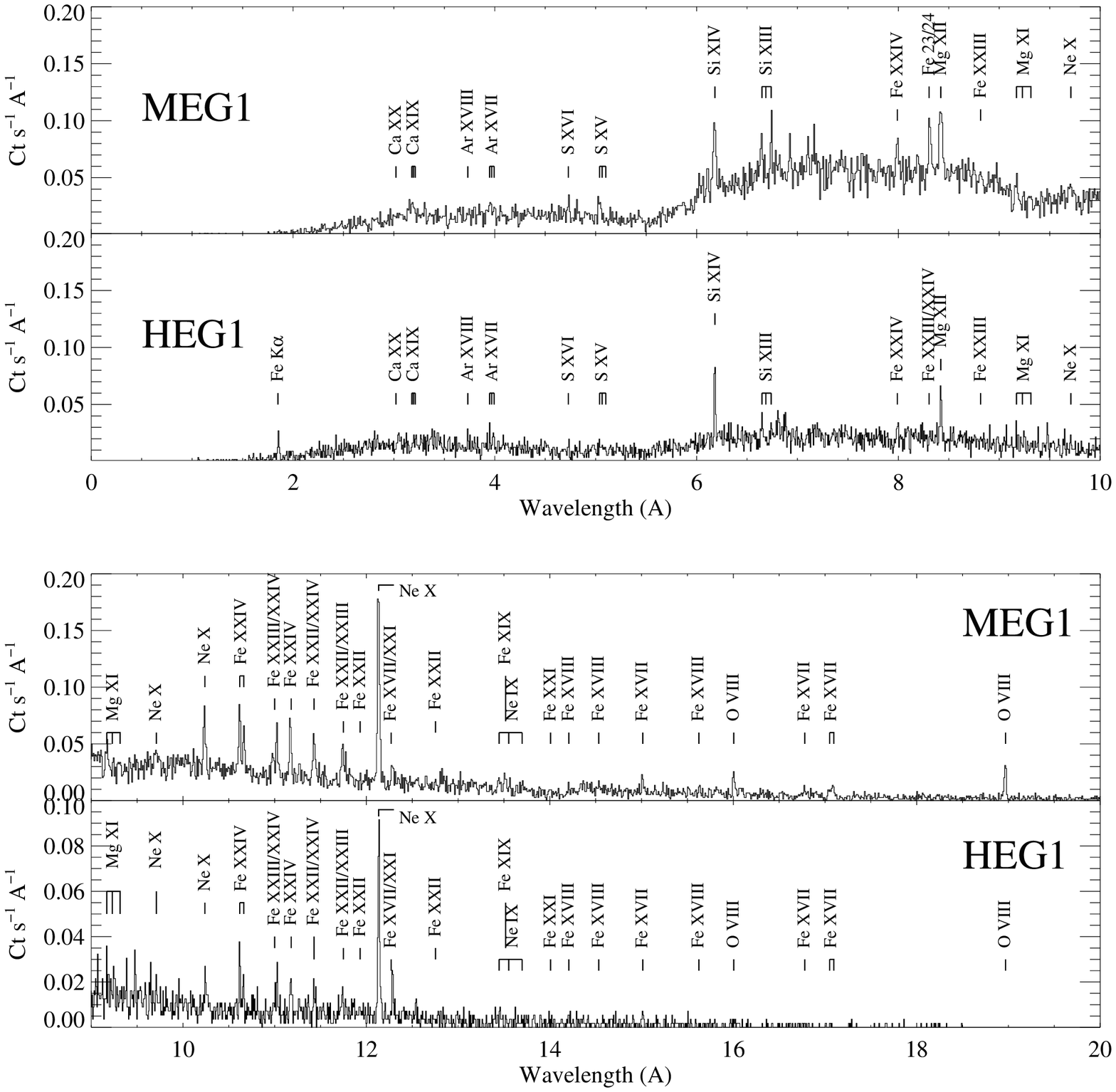}
\caption{Chandra ACIS-S/HETGS spectrum of YY Men. Several emission lines 
are labeled. Despite the strong photoelectric absorption and the small 
effective area at 25~\AA\ ($5$~cm$^{2}$), a clear N~{\sc vii} Ly$\alpha$ line 
is detected (not shown here).}
\vspace*{-3mm}
\end{figure}

We used a multi-$T$ approach and fitted the MEG and HEG data
simultaneously to obtain a discretized emission measure (EM) distribution of YY Men
along with coronal abundances. We also approached the data differently and 
extracted line fluxes to construct a continuous EM distribution and obtain coronal
abundances as well. We describe our methodology elsewhere (Audard et
al.~2004). We report here, in brief, on the 3-$T$ approach: 
a very hot (40~MK) plasma dominates the X-ray spectrum; however two other 
plasma components are detected as well (7~MK and 14~MK).
No lower $T$ component could be detected either
in the Chandra HETGS data or in the RGS spectrum. The EM ratios are 1:2:19 
(where the coolest component has EM=1), the X-ray luminosity is $L_{\rm X} =
2.2 \times 10^{32}$~erg~s$^{-1}$ (0.1--10~keV) with $N_{\rm H} = 7.5
\times 10^{20}$~cm$^{-2}$. Coronal abundances indicate a pattern
reminiscent of the inverse FIP effect observed in active RS CVn binaries
(e.g., Audard et al.~2003). However, a high N abundance (${\rm [N/H] =
+0.3}$) probably reflects a high photospheric N abundance due to CN processing 
in the stellar interior; indeed the RGS spectrum does not
show a  C~{\sc vi} Ly$\alpha$ line, indicating a low ${\rm [C/N]=-0.8}$ ratio.

\begin{figure}[!t]
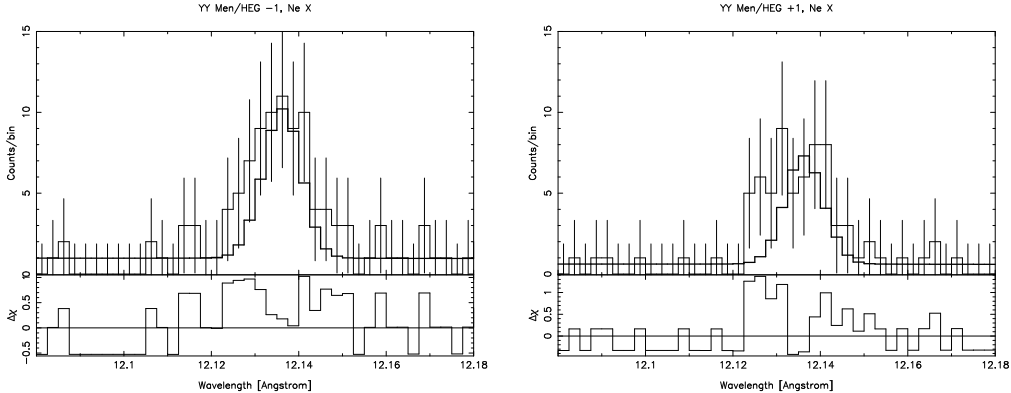

\includegraphics[angle=-90,width=0.48\textwidth]{audardfig2a.eps}\hfill
\includegraphics[angle=-90,width=0.48\textwidth]{audardfig2b.eps}
\caption{HEG $-1$ (left) and $+1$ (right) spectra of YY Men around the 
Ne~{\sc x} Ly$\alpha$ line. The instrumental profile is overlaid, emphasizing 
the line broadening (probably due to Doppler broadening). 
Similar results are {\em tentatively} reported for other lines (see text).
}
\vspace*{-3mm}
\end{figure}

Line broadening is detected in the Ne~{\sc x}
Ly$\alpha$ line ($\sigma = 7 \pm 0.7$~m\AA; Figure~2). Similar {\em tentative} 
detections are reported in Ne~{\sc x} Ly$\beta$, O~{\sc viii} 
Ly$\alpha$,  Mg~{\sc xii} Ly$\alpha$, and Si~{\sc xiv} Ly$\alpha$. 
However, S, Ar, and Ca Ly$\alpha$ lines are either too weak or at too short 
wavelength to detect line broadening. N~{\sc vii} Ly$\alpha$, however, does 
{\em not} show evidence of broadening. Th Ne~{\sc x} line broadening is unlikely
to be due to rotation since (i) it would require $v \sin i \gg 50$~km~s$^{-1}$,
and (ii) it would be easily detected in the longer wavelength lines.
Thus we interpret the observed broadening 
as Doppler thermal broadening in the very hot corona of YY Men. The N~{\sc vii}
line, despite the lower atomic mass of N than of Ne, is probably not broadened
because it is mostly formed by the lower $T$ components, while the Ne lines are 
mostly formed by the very hot plasma (as are the lines with tentative detections
of line broadening). 
This is also supported by the absence of line broadening in Fe lines, probably 
due to the larger Fe atomic mass. The observed Doppler width ($\sigma_{\rm D} =
(\lambda / c) \times 
\sqrt{kT/m}$) corresponds to a plasma $T$ in the range 60--90~MK. Possibly
additional broadenings, e.g., due to rotation, contribute to the total line
width, which would lower the $T$ range. 
 
\vspace*{-4mm}
\section{V824 Arae = HD 155555}
The nearby (31~pc) binary consists of G5~V+K0~V stars tidally locked 
with an orbital period of 1.68~d. Based on its kinematics and its high Li 
abundance, V824 Ara is either in the pre-main-sequence phase 
(Pasquini et al.~1991) or in the zero-age-main-sequence phase (Mart\'\i n \& 
Brandner 1995), with an age of $10^{7}-10^8$~yr. Its common proper-motion 
dM4.5e companion (33$^{\prime\prime}$ away), LDS 587B, also displays high
chromospheric and coronal 
activity. Since V824 Ara is so young, its photosphere should have 
near-solar photospheric composition. 

Figure~3 shows the Chandra ACIS-S/HETG MEG spectrum of V824 Ara together
with a best-fit 4-$T$ plasma model. The EM distribution differs from YY Men; it 
is flatter with $T$ of 3.7, 7.7, 14, and 28~MK, EM ratios of 
1:3:2:2.7, and $L_{\rm X}=  4.4 \times 10^{30}$~erg~s$^{-1}$ (0.1--10~keV).
The coronal abundances show a distinct inverse FIP effect (${\rm [Fe/H] =
-0.5}$, ${\rm [O/H] = -0.3}$, ${\rm [Ne/H] = -0.1}$), except at very low FIP 
($<7$~eV), where the Al and Ca abundances seem to increase. A similar
pattern was observed in the coronal abundances of the RS CVn binary $\sigma^2$
CrB (Osten et al.~2003). The LDS 587B companion was sufficiently bright
($L_{\rm X}=  2.8 \times 10^{29}$~erg~s$^{-1}$) and spatially resolved to 
produce a distinct grating
spectrum that reveals a similar $T$ structure (although EM ratios for
a 3-$T$ fit are 1:1.1:1.6). Preliminary coronal abundances indicate no obvious
correlation with the FIP.

\begin{figure}
\includegraphics[angle=-90,width=0.45\textwidth]{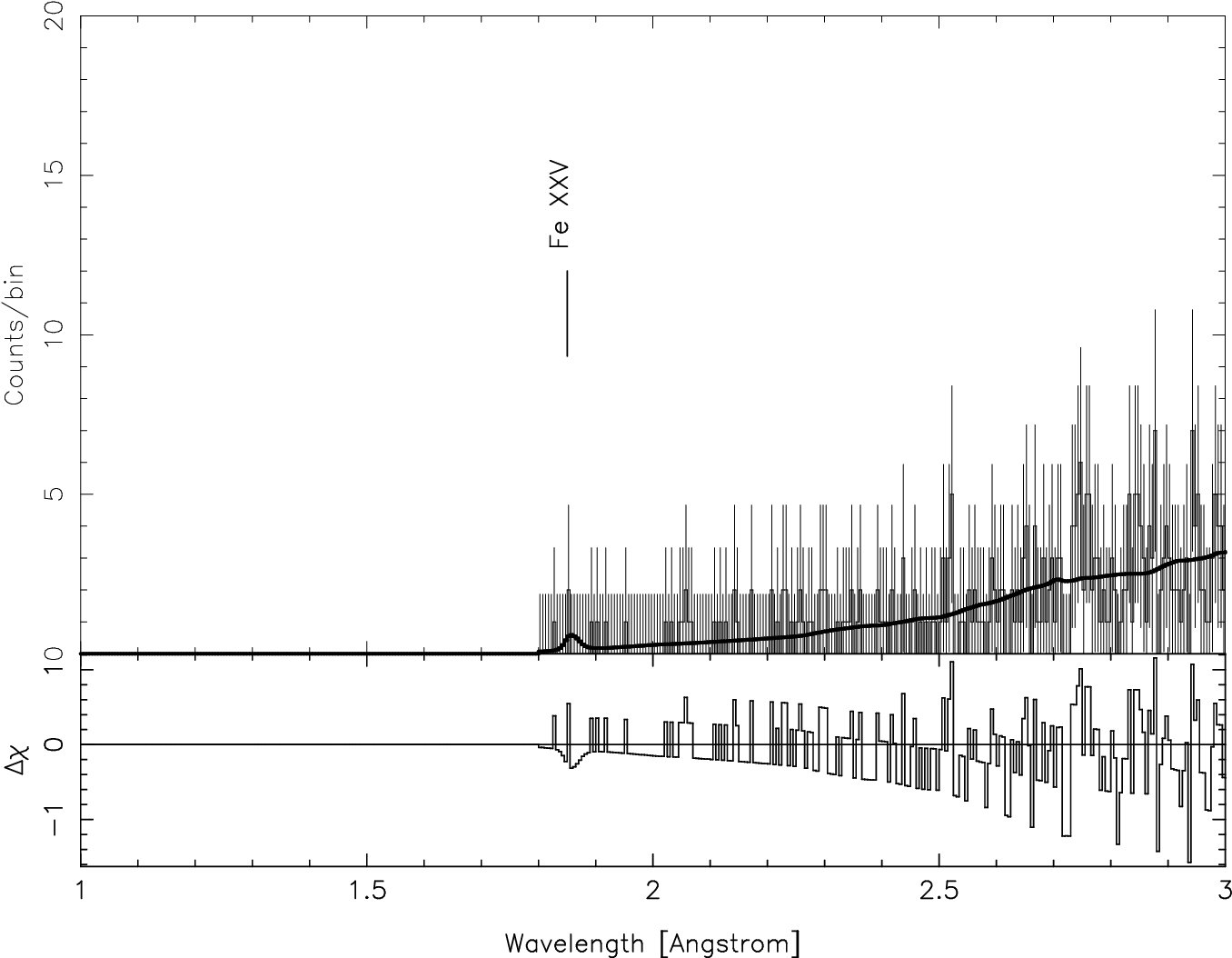}\hfill
\includegraphics[angle=-90,width=0.45\textwidth]{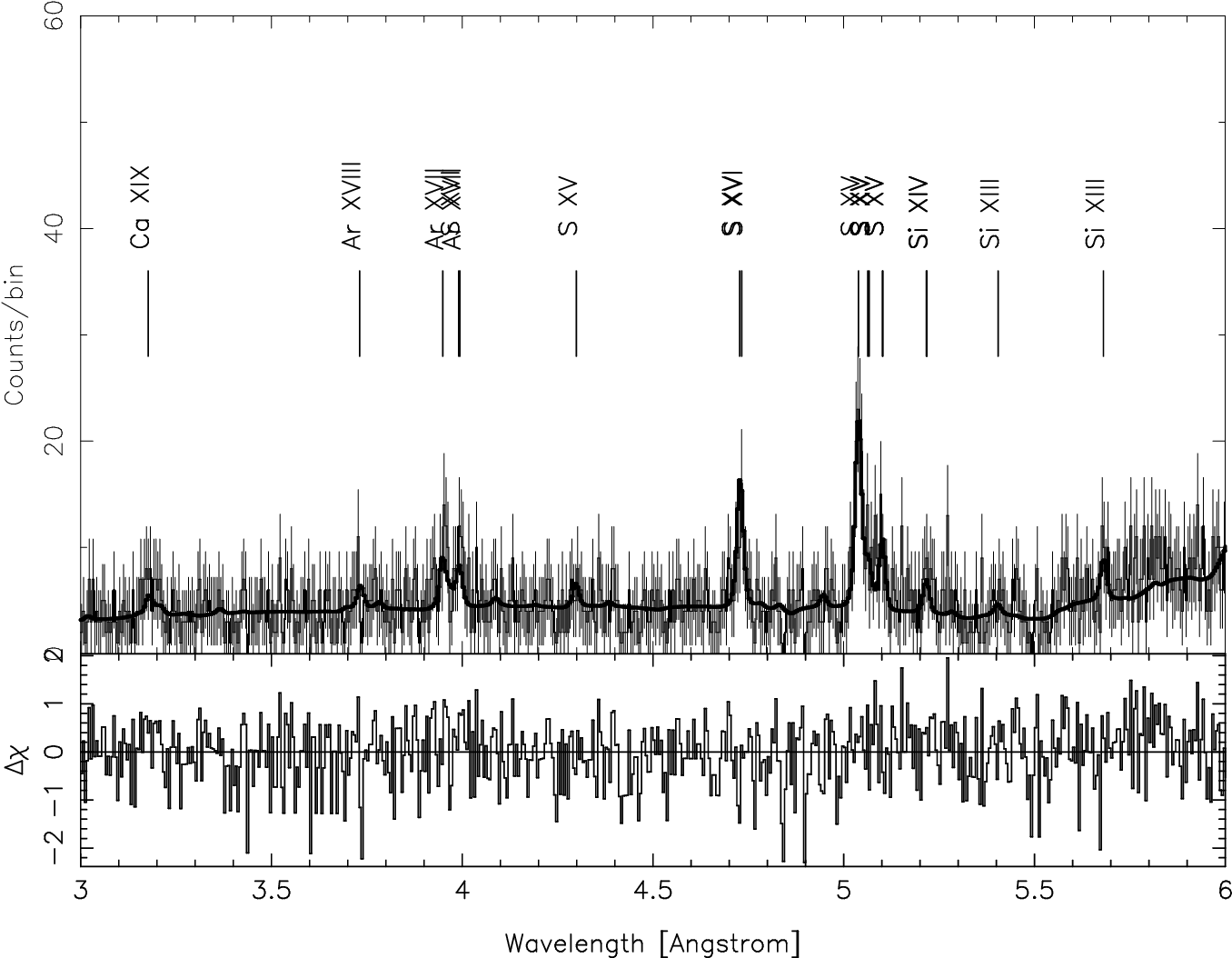}\\
\includegraphics[angle=-90,width=0.45\textwidth]{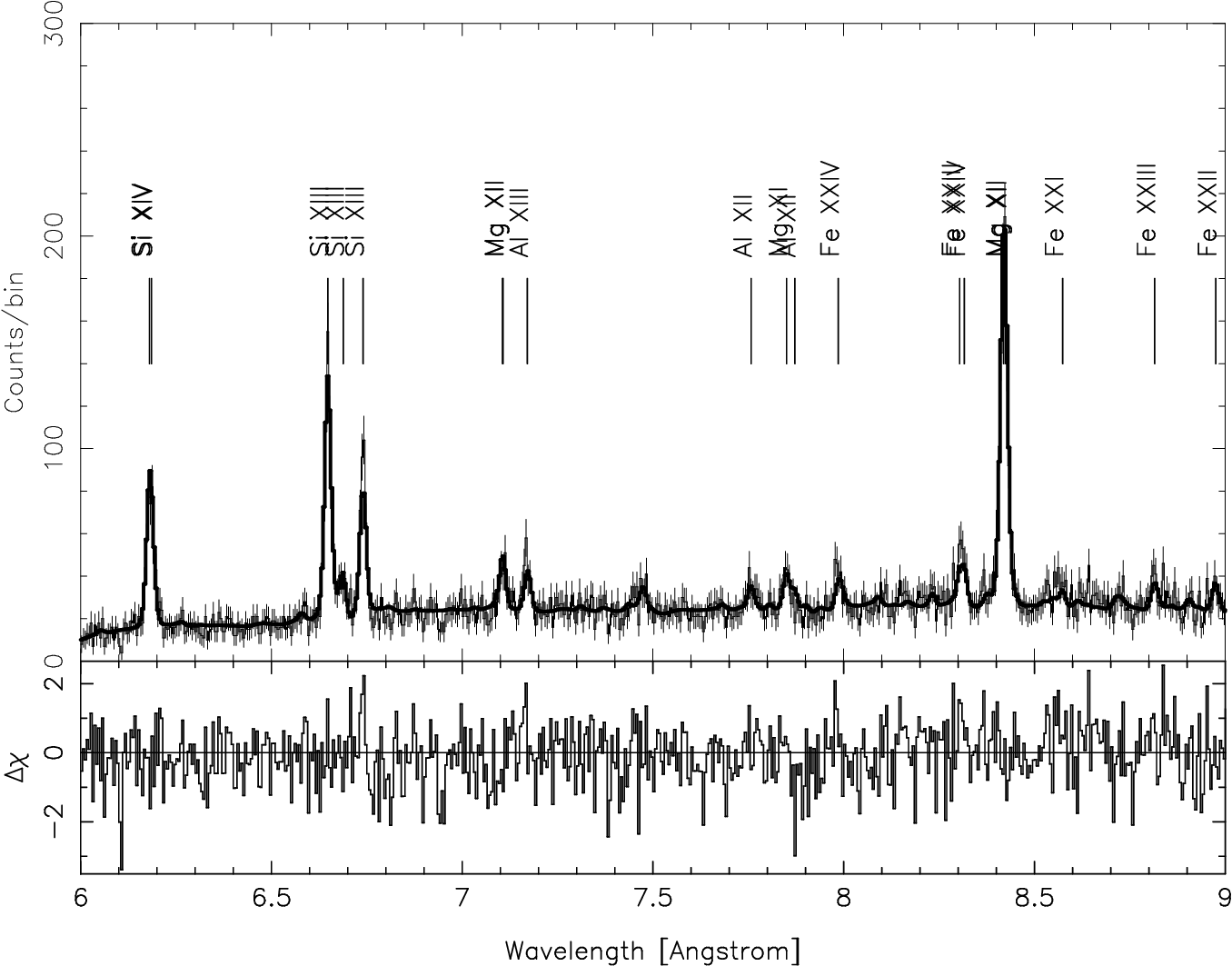}\hfill
\includegraphics[angle=-90,width=0.45\textwidth]{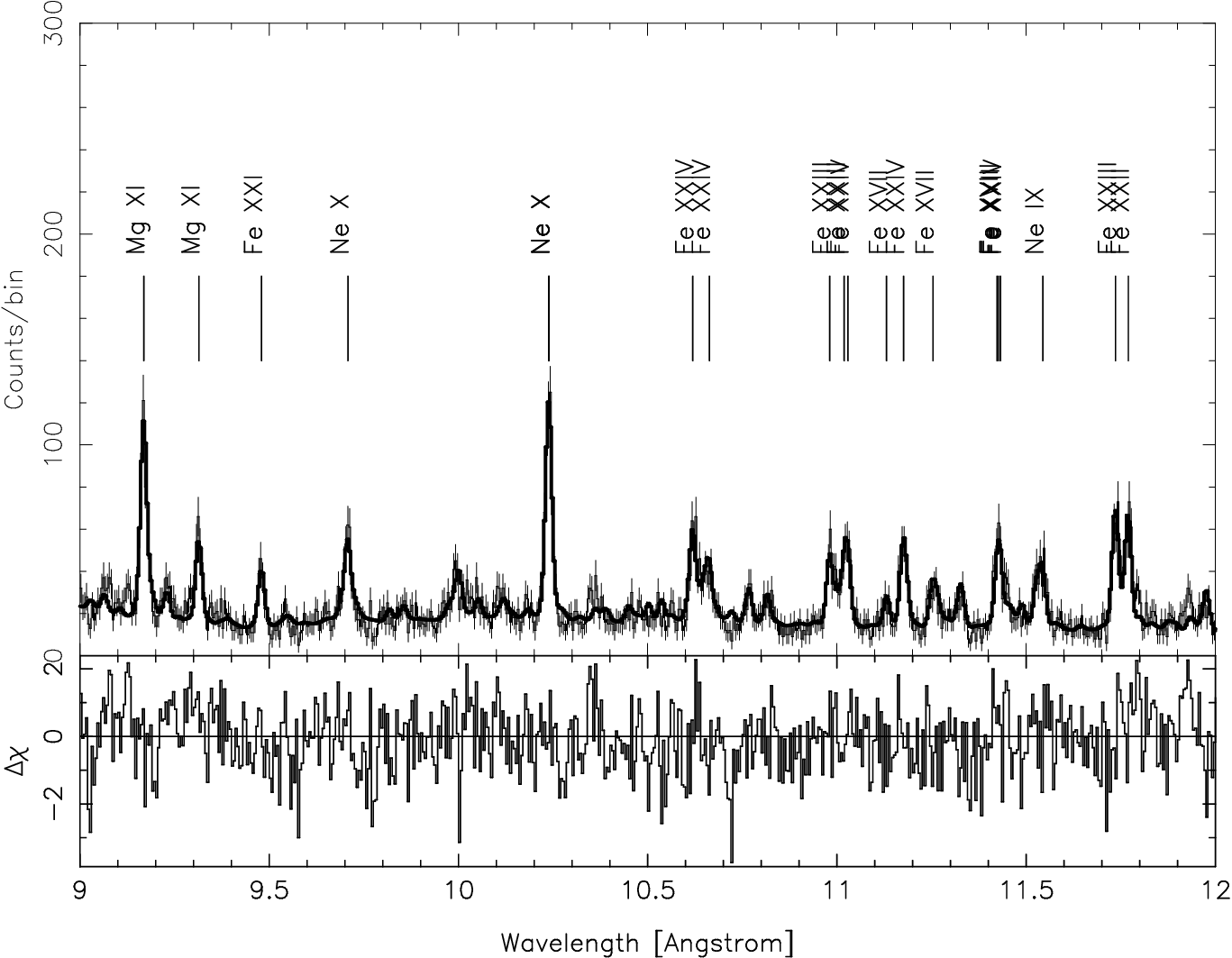}\\
\includegraphics[angle=-90,width=0.45\textwidth]{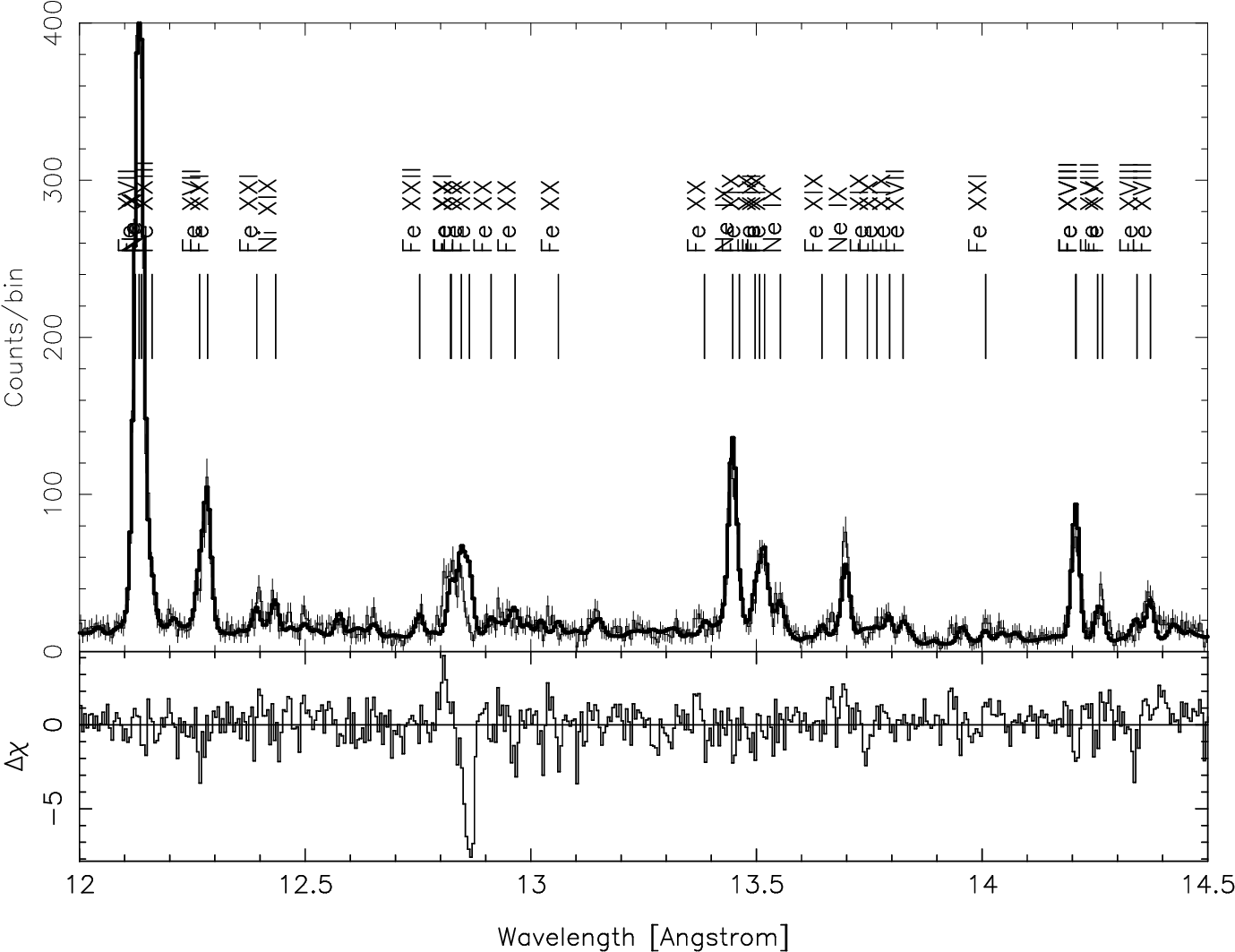}\hfill
\includegraphics[angle=-90,width=0.45\textwidth]{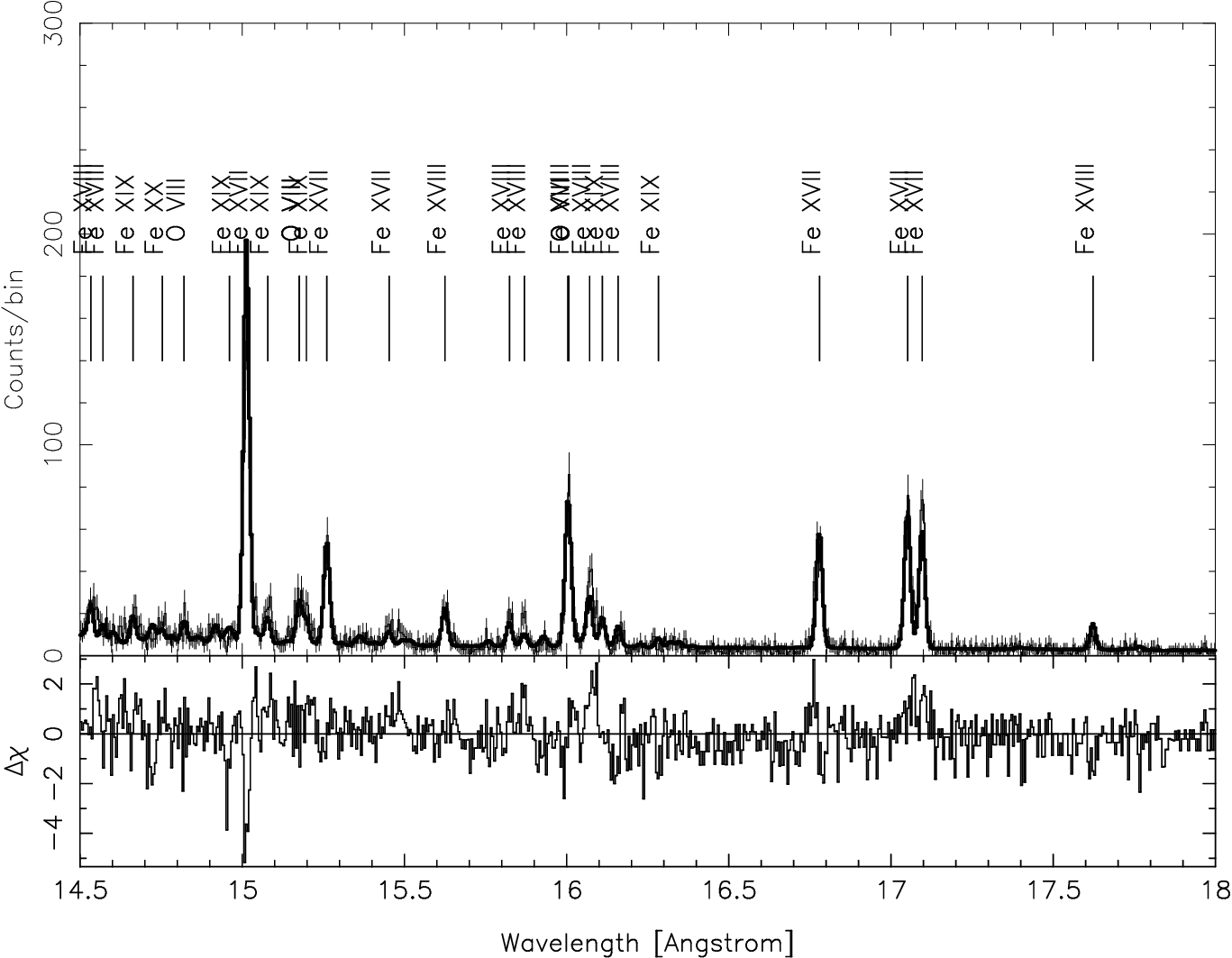}\\
\includegraphics[angle=-90,width=0.45\textwidth]{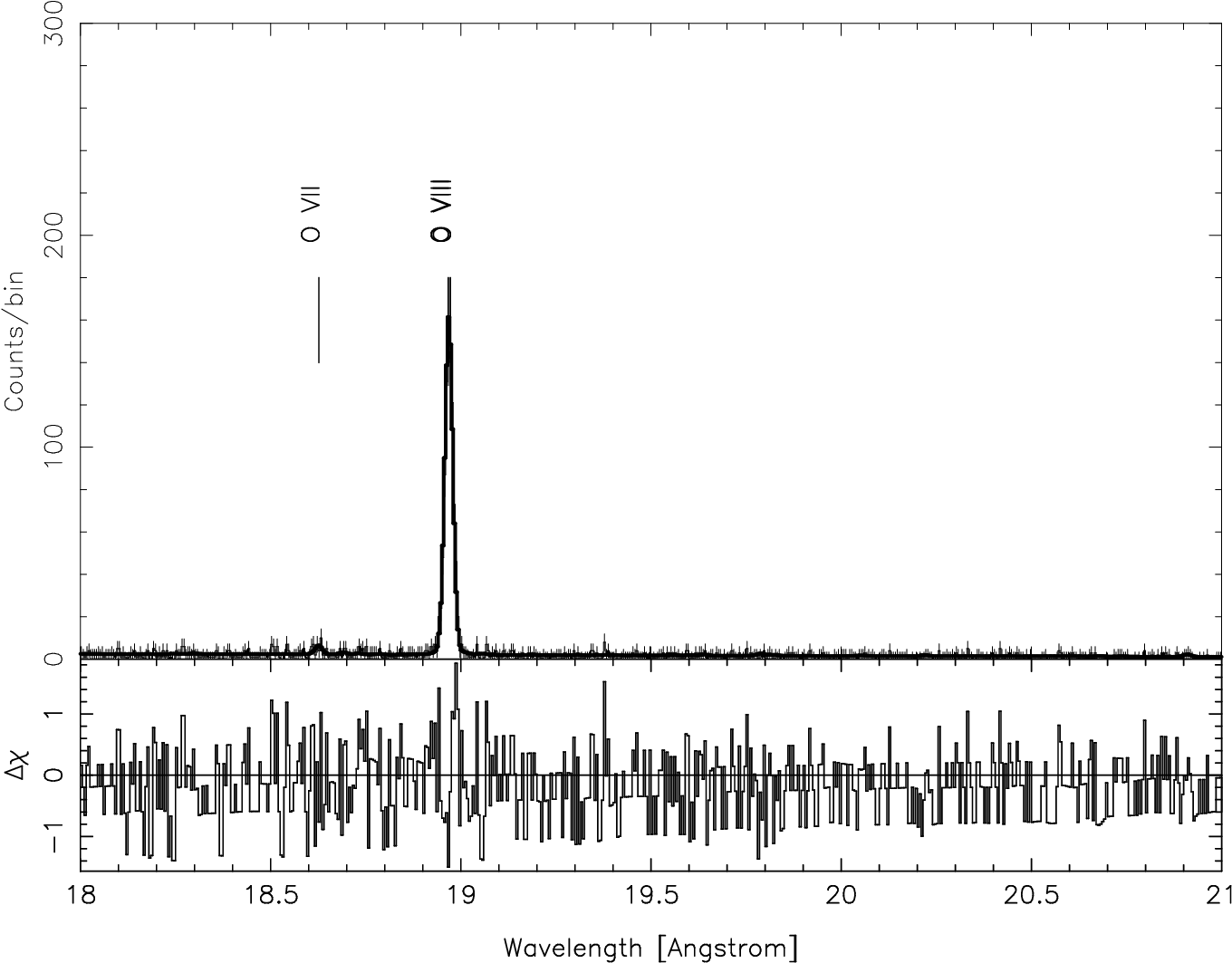}\hfill
\includegraphics[angle=-90,width=0.45\textwidth]{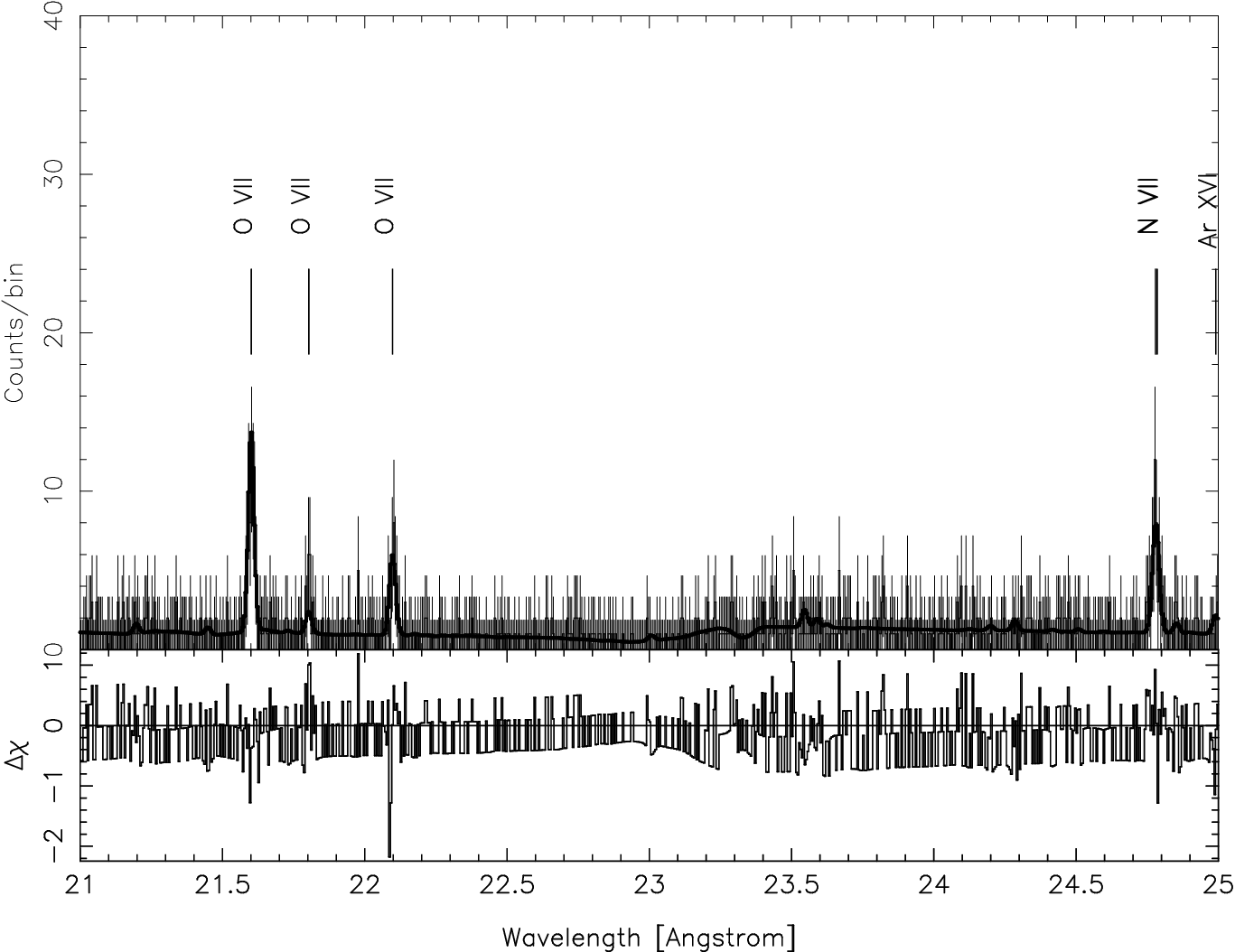}
\caption{Chandra ACIS-S/HETG spectrum of V824 Ara with major emission
lines labeled. A 4-$T$ best-fit model is overlaid with a thick line. Residuals
are shown in the lower panels. Note the bad fit of the Fe~{\sc xx} blend at
12.8~\AA\  which can be explained by inaccurate wavelengths estimated 
from laboratory measurements (from Brown et al.~2002, see their Fig.~3b) and 
listed in the APEC~1.3 database.}
\end{figure}

Plasma densities can be derived from line ratios, typically from the forbidden
and intercombination lines of He-like transitions. The well-developed HETGS
spectrum of V824 Ara reveals no variation from the low density limit in the
various He-like triplets, suggesting that the dominant X-ray emitting plasma has
an electron density below $10^{10}$~cm$^{-3}$. Since coronae are
inhomogeneous, this result does not mean that no high-density plasma exists.
However, it suggests that such plasma has a very low volume 
(since $EM \sim n^2_e V$). 

\vspace*{-4mm}
\section{Conclusions}

We have presented results of observations of YY Men and V824 Ara with Chandra and
XMM-Newton. The highly active stars show contrasting X-ray spectra: the EM
distribution of YY Men is dominated by a very hot (40~MK) plasma but has
measurable EM between 6 and 15~MK, whereas the binary V824 Ara displays a rather
flat EM distribution from 3~MK to 30~MK. Despite their differences, their
coronal abundances (relative to the solar photospheric composition) generally
follow the inverse FIP effect observed in other active binaries. However,
coronal abundances in the M dwarf LDS 587B do not correlate
with the FIP. The coronal N abundance in YY Men is enhanced (C is depleted),
probably reflecting a true photospheric N/C enhancement due to mixing of 
CN-cycle material in the stellar interior. YY Men also shows 
line broadening in Ne~{\sc x} Ly$\alpha$ and possibly in other 
lines, which we interpret as Doppler broadening.

The general picture is that coronal abundances in very active stars display 
an inverse FIP effect, while inactive stars show a solar-like FIP effect
(e.g., G\"udel et al.~2002). However, most current studies are limited to comparing 
{\em stellar} coronal abundances to {\em solar} photospheric abundances 
because of the lack of accurate photospheric measurements in active stars.
Thus caution is in effect, since coronal abundances could also be affected by 
anomalous photospheric compositions as shown here for YY Men, apart from 
selective, FIP-dependent coronal enrichment. Improvements in the photospheric 
composition in stars will help better understand the coronal composition.

\acknowledgements
We acknowledge support from the Swiss National Science Foundation (fellowship 81EZ-67388 \& grant 
2000-058827), from NASA to Columbia University for XMM-Newton mission 
support and data analysis, and from SAO grant \mbox{G02-3016X}.


\end{document}